\documentclass[a4paper,fleqn,usenatbib]{mnras}
\usepackage{amssymb,amsmath,graphicx,psfrag,mathtools}
\usepackage[T1]{fontenc} %MNRAS
\usepackage{ae,aecompl} %MNRAS
\usepackage{psfrag}
\psfrag{Sun}{$\odot$}
\usepackage{url}
\title[Arrokoth's Necklace]{Arrokoth's Necklace}
\author[J. I. Katz \& S. Wang]{
	J. I. Katz$^{1}$\thanks{E-mail: katz@wuphys.wustl.edu}
	and S. Wang$^{2}$%MNRAS
\\
$^{1}$Department of Physics and McDonnell Center for the Space Sciences,
Washington University, St. Louis, Mo. 63130 USA \\%MNRAS
$^{2}$Department of Physics, Washington University, St. Louis, Mo. 63130 USA
}
\date{Accepted XXX.  Received YYY; in original form ZZZ} %MNRAS
\pubyear{2020} %MNRAS
\date{\today}
\begin{document} %MNRAS
\label{firstpage} %MNRAS
\pagerange{\pageref{firstpage}--\pageref{lastpage}} %MNRAS
\maketitle %MNRAS
\begin{abstract}
	Flyby images of (486958) Arrokoth (Ultima Thule, 2014 MU$_{69}$)
	show a comparatively bright ``necklace'' in the neck, or cleft,
	between its two lobes, in contrast to its generally low albedo.
	{We suggest that the necklace may be the result of thermally
	controlled ice deposition.}  The necklace is found in the most
	(orbitally averaged) shaded part of the surface.  It may consist of
	clean, high albedo, ice condensed from vapor sublimed by dirty, low
	albedo, ice elsewhere; ice accumulates where the {\it maximum\/}
	temperatures are the lowest.  Ammonia and propane have the necessary
	{mesovolatile} vapor pressure.  Surrounding gas in the
	proto-Solar System would facilitate redeposition of molecules
	sublimed by warmer parts of the surface into the cleft, as well as
	smoothing the surface and explaining, by hydrodynamic drag,
	Arrokoth's slow (compared to its breakup rate) rotation. 
	{Alternatively, a layer of hoarfrost thick enough ($\gtrsim 0.1\,
	\mu$) to have a high albedo could have formed more recently.}
\end{abstract}
 %MNRAS
%\maketitle
\section{Introduction}
(486958) Arrokoth (Ultima Thule, 2014 MU$_{69}$) \citep{S19,G20,McK20,S20},
a cold classical Kuiper belt object, became the most distant object visited
by a spacecraft following the flyby by New Horizons January 1, 2019.
Imaging \citep{McK20} revealed it to be a contact binary consisting of two 
lobes with equivalent diameters of about 19 and 14 km, rotating
with a period of 15.92 h.  Its location in the outer Solar System with
orbital semi-major axis $a = 44.6\,$AU and eccentricity $e = 0.042$
\citep{P18,S19} suggests a predominantly icy composition, although its low
{(compared to those of pure ices)} albedo $\approx 0.06$ \citep{Ho21}
implies some mineral (``dirt'') admixture with the
mineral matter frozen into a continuous icy matrix\footnote{If there were
voids between particles of clean ice Arrokoth would resemble snow that has
high albedo as a result of scattering of light at the surfaces of the
transparent ice particles, unless mineral matter constitutes a substantial
fraction of its volume.  Comparison of the observed albedo to that of the
presumptive mineral content may constrain its microstructure: very small
($\ll \lambda$, the wavelength of visible light) mineral particles frozen
into the ice without voids would produce a dirty ice albedo less than that
of macroscopic solids. This may be demonstrated by adding water to granular
sugar with about $10^{-4}$ admixture of carbon black:  Without water, the
mixture is light grey, but adding water reduces the discontinuity in
refractive index at the sugar grains, producing a black slurry.  A mixture
of ices and larger mineral particles, even without voids, has an albedo
larger than that of bulk mineral because of scattering from the surface of
the ice.} or a surface of photochemically processed tholins.

The most striking feature of images of Arrokoth is a bright ring or
``necklace'' where the two lobes are in contact.  This paper proposes that
the necklace is produced by thermal evaporation of ice, likely ammonia ice,
from the most strongly heated portions of the surface and the deposition of
some fraction of it on the neck, the most shaded and coolest portion.
Deposition from the vapor would produce a layer of nearly pure solid ice
with high albedo.  Brighter regions in apparent craters and other surface
depressions may have a similar origin.  {\citet{L20,St20}}
have discussed the presence of {hyper- and refractory} volatiles in
Arrokoth and their effects on its {bulk} evolution {and spin-down},
but have not addressed the question of {moderately volatile, metastable
or mesostable ices and} its bright ``necklace''.  {The neck is a low of
both {\it peak\/} temperatures, because it is shaded from sunlight at most
times, and of gravitational potential, because it is near the center of mass.
Material may be deposited there by condensation or by gravitational flow;
this paper is concerned with the former hypothesis.} 

The peculiar shape of Arrokoth, consisting of two smooth lobes in contact
with a narrow cleft between them, rotating well below their break-up rate,
also requires explanation: Why did the process that smoothed the lobes not
fill in the cleft?  This paper proposes that the lobes formed and smoothed
in the gas-filled proto-Solar System.  Bodies immersed in gas that prevents
the free escape of sublimed material smooth by evaporation from the convex
and warmest parts of their Solar-heated surfaces and recondensation on
cooler, shadowed, concave parts.  Drag {from immersing gas} also slows
the rotation of a contact binary, {possibly explaining the shedding of
angular momentum} { required to produce the slowly spinning bilobate object
seen today \citep{Ly20,McK20,St20}}.
\section{Heat Flow}
\subsection{Steady State}
\label{steadystate}
The steady state daytime temperature of a grey (equal albedos and 
emissivities for Solar and thermal radiation) surface, considering only
local radiation absorption and emission, is
\begin{equation}
	\label{Tsteady}
	T_{steady} = (I_\odot\sin{\theta}/\sigma_{SB})^{1/4},
\end{equation}
where $I_{\odot}$ is the Solar intensity, $0 \le \theta \le \pi/2$ the
grazing angle of the Sun's rays to the surface and $\sigma_{SB}$ the
Stefan-Boltzmann constant.  For $\theta = \pi/2$ (Sun at the zenith) this
varies around the orbit from 58.4 K to 60.6 K.

In fact, these extreme temperatures are not achieved because of conduction
to and from the interior of the body.  The deep interior of a grey sphere
comes to the mean temperature
\begin{equation}
	\label{Tav}
	\langle T \rangle = (I_\odot/4\sigma_{SB})^{1/4} \approx 42\,
	\text{K}.
\end{equation}

The grey-body temperature can only be an approximation to the steady state
surface temperature.  Actual steady state temperatures are a factor
$[(1-A_\odot)/(1-A_{thermal})]^{1/4}$ times the grey body values, where
$A_\odot \approx 0.06$ \citep{S19,Ho21} is the spectral-, polarization- and
angular-averaged bolometric Solar albedo (insufficient information exists to
determine this precisely) and $1-A_{thermal}$ is the emissivity, similarly
averaged over the Planck function at the surface temperature.  For {pure,
clean} NH$_3$ ice at relevant temperatures $1-A_{thermal} \approx 0.77$
(Appendix), possibly increased by the presence of mineral ``dirt'' or a
surface layer of tholins.  Steady state temperatures of clean NH$_3$ ice are
therefore expected to be about 5\% higher than those of a grey body, a
correction that is not included explicitly here because it is not large and
because of its uncertainty.
\subsection{Transients}
The surface of Arrokoth is not in thermal steady state.  Several time scales
enter: the short ($P_{rot} = 15.92$ h) rotation period, the longer ($P_{orb}
= 298$ y) orbital period, the thermal conduction time scale of the body as a
whole ($t_{deep} \sim 3 \times 10^5$ y) and the evolution time scale
($t_\odot \sim 5 \times 10^9$ y) of the Solar luminosity, which is
approximately the age of Arrokoth.  The rotation axis is only $9^\circ$ from
the orbital plane \citep{G20}, so that most portions of the surface are
continuously sunlit and then continuously dark, each for a substantial
fraction of the 298 y orbital period.  This produces temperature excursions
about the mean $\langle T \rangle$.

There is yet another characteristic time scale.  As the insolation varies
with the rotational and orbital cycles, Solar heating and thermal radiation
cooling of the surface compete with conductive heat flow into or out of the
interior as sinks or sources of energy.  In a time $\Delta t$ the thermal
diffusion wave penetrates a distance
\begin{equation}
	\label{Deltax}
	\Delta x \approx \sqrt{{\cal K} \Delta t/(C_{p,s}\rho)},
\end{equation}
where $C_{p,s}$ is the heat capacity (per unit mass) of the solid, $\rho$
its density and $\cal K$ its thermal conductivity, carrying an energy per
unit area $C_{p,s} \rho \Delta T \Delta x$, while the surface absorbs and
emits $\sim I_\odot \Delta t$.  Equating these energies defines the
characteristic thermal relaxation time of the surface
\begin{equation}
	\label{relax}
	t_{relax} \equiv {(\Delta T)^2 C_{p,s} \rho {\cal K} \over I_\odot^2}
	\approx 1.5 \times 10^9\,\text{s} \approx 50\,\text{y}
\end{equation}
and
\begin{equation}
	\Delta x \approx {{\cal K} \Delta T \over I_\odot} \approx 130\,
	\text{m},
\end{equation}
where we have taken the heat capacity and thermal conductivity to be those
of solid NH$_3$ \citep{PMB71,KMK68}.  Most other ices (water ice \citep{GS36}
and CO$_2$ ice \citep{GE37}) have similar values, {while the thermal
conductivity of CH$_3$OH is about 30 times less \citep{KKSR09} and its
$t_{relax}$ and $\Delta x$ are correspondingly less.  Changes in irradiation
on shorter time scales have little effect on the surface temperature, while
on time scales (such as the orbital period) between $t_{relax}$ and
$t_{deep}$ the surface temperature relaxes to the mean insolation.}

\begin{figure}
	\centering
	\includegraphics[width=0.95\columnwidth]{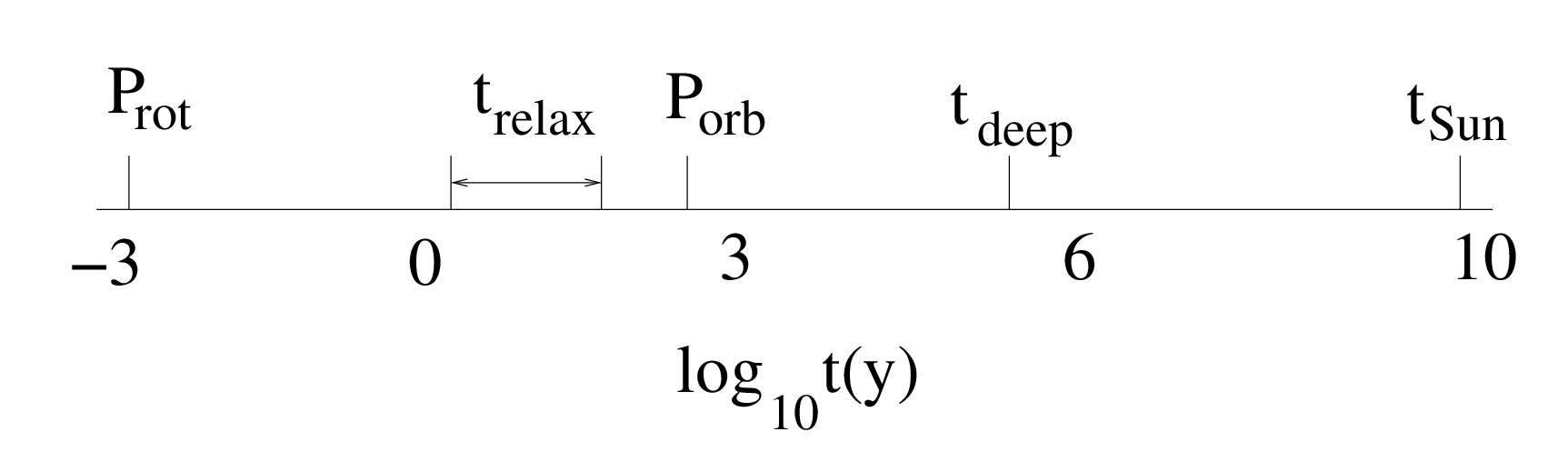}
	\caption{\label{tscale} Characteristic time scales.  The range
	for $t_{relax}$ corresponds to the range of possible compositions,
	with shorter values for low-conductivity materials like glassy
	CH$_3$OH or snows, and longer values for water ice.}
\end{figure}

{The several time scales are ordered
\begin{equation}
	P_{rot} \ll t_{relax} \lesssim P_{orb} \ll t_{deep} \ll t_\odot.
\end{equation}
{These inequalities are illustrated in Fig.~\ref{tscale}.} The strong
inequalities simplify the analysis by separating the time scales on which
some processes are insignificant while others have gone to completion.  The
rotational slowing time $t_{slow}$ is not included in these inequalities
because it increases by many orders of magnitude as the protoplanetary
nebula dissipates, so that at present it far exceeds $t_\odot$, while during
the period of most efficient (diffusive) vapor transfer from warmer regions
to the neck (Sec.~\ref{sgas}) it is less than $t_\odot$, although not by
orders of magnitude.}

The rotation axis is only $9^\circ$ from the orbital plane \citep{G20}.  On
time scales $\gg t_{relax}$ thermal conduction is unimportant and surface
temperatures relax to the steady state values corresponding to their mean
insolation.  On shorter time scales conduction is important, the internal
specific heat acts as thermal ballast, and temperature varies less than it
would on longer time scales.  Rotational modulation of insolation produces
only small variation of surface temperature; the surface temperature is
nearly constant at the value corresponding to the rotationally averaged
insolation (which, in general, is not the deep interior $\langle T
\rangle$).  Relaxation to the radiative equilibrium temperature is, at best,
approximate for the orbital (seasonal) cycle.

For a slab of thickness $r$ (a fair approximation {to the highly flattened
shape of Arrokoth's lobes}) with radiatively heated temperature $T$ that
varies by $\Delta T$ across its surface, the ratio of conductive heat flow
from warmer parts of the surface through the deep interior to its black body
radiation is, {\it in steady state\/},
\begin{equation}
	\label{Fcond}
	{F_{cond} \over F_{rad}} \sim {{\cal K} \Delta T \over r
	\sigma_{SB} T^4} \sim 2 \times 10^{-2},
\end{equation}
where at $T = 40\text{--}60\,$K the thermal conductivity of water ice, the
likely dominant component of Arrokoth, ${\cal K} \approx 1 \times
10^6\,$erg/(cm-s-K) \citep{S80} and we take $T = 60\,$K, $\Delta T = T_{max} -
\langle T \rangle = 18\,$K and $r = 10\,$km.  $\cal K$ of NH$_3$ ice is
about half as large \citep{KMK68}, {while $\cal K$ of methanol glass,
the observed surface, but possibly very shallow, constituent, is about $2
\times 10^4$ erg/(cm-s-K) \citep{KKSR09}.}  Thermal conduction into and from
the deep interior is a minor contributor to the thermal balance of surfaces
exposed to sunlight {for times $\gg t_{relax}$ while conduction from a
layer of thickness $\sim \Delta x$ averages the temperature over a time
$\sim t_{relax}$.}

{In contrast, a surface long exposed to dark space is heated only by
conduction through the interior and approaches a steady state temperature
$T_{dark}$ given by
\begin{equation}
	\sigma_{SB} T_{dark}^4 = {{\cal K} (\langle T \rangle - T_{dark})
	\over r}.
\end{equation}
The resulting $T_{dark} \approx 23\,$K is consistent with the measured
\citep{G20,U20} dark side temperature of Arrokoth of $29 \pm 5\,$K.}
However, in superficial layers of thickness $\sim \Delta x$
(Eq.~\ref{Deltax}) thermal conduction dominates radiative processes on the
time scale $P_{rot}$ and is significant on the time scale $t_{relax}$
(Eq.~\ref{relax}).

Because $t_{deep} \gg P_{orb}$, conduction into the deep interior does not
significantly affect orbital surface temperature variations; this is implied
by Eq.~\ref{Fcond}.  The relevant sublimation rates are very low so the
contribution of latent heat to the energy balance is negligible.

%This condition is at most barely met.  The actual peak surface temperature
%is $\approx 50\,$K, between $T_{max}$ and $\langle T \rangle$, and depends
%on unknown details of the shape and orientation of Arrokoth in addition
%to its infrared emissivity.
The maximum surface temperature is approximately (only approximately,
because $P_{orb}$ does not greatly exceed $t_{relax}$) given by
Eq.~\ref{Tsteady} with $\theta = \pi/2$.  This occurs near the pole of the
summer hemisphere at solstice {({\it i.e.\/}, the sunlit surface as seen in
the 2019 New Horizons flyby)}.  At this time the cleft between the lobes,
the subject of this paper, is in deep shadow and its temperature approaches
$\langle T \rangle$ (Eq.~\ref{Tav}) as a result of heat conduction from the
interior.  The cleft, receiving very little sunlight {averaged over an
orbit}, is a thermal probe of the deep interior.

\begin{figure}
	\centering
	\includegraphics[width=0.9\columnwidth]{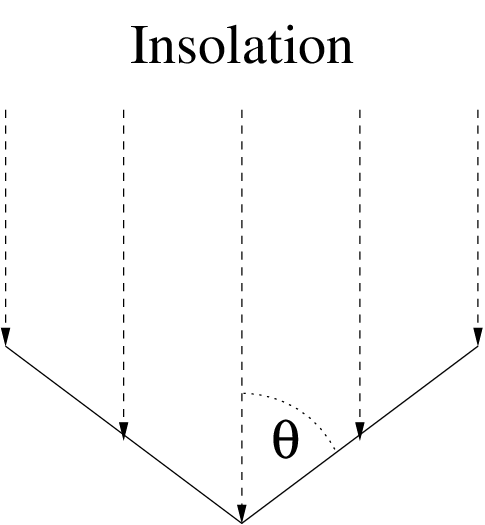}
	\caption{\label{cleft} Insolation of a wedge-shaped cleft between
	the lobes at equinoctal noon.  This idealized geometry can only be
	an approximation to the actual geometry \citep{G20}, but illustrates
	its essential features.}
\end{figure}

At equinoxes, the illumination of an ideal wedge-shaped cleft with the Sun
aligned with its midplane is shown in Fig.~\ref{cleft}.  In this geometry 
the two sides of the cleft come to the same temperature so that radiative
exchange between them produces no net heating or cooling; in more complex
geometry they come to an average temperature.  Averaging over the rotation
leads to a mean insolation $I_\odot \sin{\theta}/\pi$.  The thermal emission
of the cleft and the sunlight it absorbs are both proportional to the area
of its mouth so that its {\it steady state\/} temperature
\begin{equation}
	\label{Tcleft}
	T_{cleft} = \left({4 \over \pi}\right)^{1/4} \langle T \rangle =
	1.062 \langle T \rangle
\end{equation}
is independent of $\theta$ and its width (or depth).

The cleft is illuminated for only a fraction $\sim \theta/\pi$ of the
orbit, a time $\sim \theta P_{orb}/\pi \ll t_{relax}$, so this brief period
of radiative heating has little effect on its surface temperature.  Away
from the equinoxes, the cleft is in darkness, without Solar heating, and is
a probe into the deep interior, warmed only by heat conduction.  Its surface
temperature remains closer to $\langle T \rangle$ than does the temperature
of any other part of the surface; it has the lowest {\it maximum\/} and the
highest {\it minimum\/} temperature because of its close thermal coupling to
the interior.  Because of the extreme temperature sensitivity of vapor
pressures, accumulation of volatiles is determined by evaporation at the
maximum temperature a surface element reaches (evaporation is negligible at
any significantly lower temperature, almost regardless of how much lower;
Sec.~\ref{net} and Eq.~\ref{DTmin}), and sublimed volatiles accumulate in
the cleft.

At earlier times in the main sequence history of the Sun it was less
luminous, so that Arrokoth was colder.  Even earlier, the proto-Sun on its
Hayashi track \citep{KWW12} was {orders of magnitude} more luminous than it
is today, so that if Arrokoth had formed then it might have been
significantly warmer than it is today.  This is complicated by the
likelihood that dust in the proto-{planetary disc midplane \citep{K97,N20}}
made it significantly opaque, reducing the temperature of objects like
Arrokoth immersed in it.  {A further complication is the possibility of
heating by short-lived radioactivities {\citep{L20}}}.
\section{Vapor Pressures}
\begin{figure}
	\centering
	\includegraphics[width=3in]{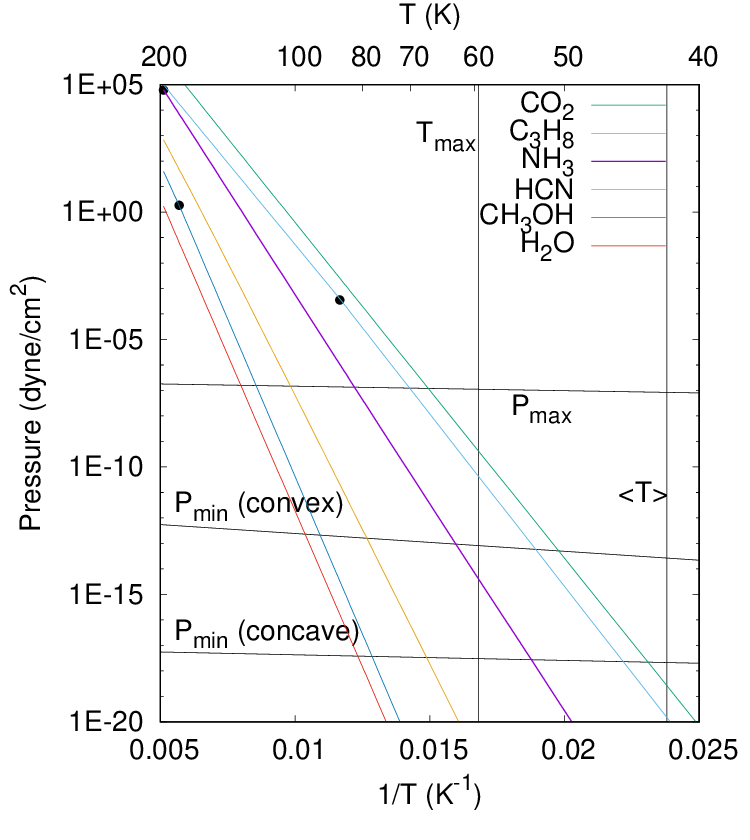}
	\caption{\label{Pvap} Vapor pressures { {\it vs.\/} temperature
	for various possible major component ices of Arrokoth}.  Curves are
	in the same order, top to bottom, as the key.  The vertical lines
	indicate $T_{max} = 59.5\,$K, the mean (over the orbit) grey-body
	sub-solar (maximum) steady-state temperature and $\langle T \rangle
	= 42\,$K, the deep interior temperature and temperature of a
	spherical body of uniform temperature.  Vapor pressures above
	$P_{max}$ would lead to evaporation of an object the size of
	Arrokoth in the age of the Solar System.  Vapor pressures below the
	geometry-dependent $P_{min}$ could not produce a layer of high
	albedo pure ices; these vapor pressures are shown for NH$_3$ but
	depend slightly on molecular weight through the thermal velocity
	(Eq.~\ref{recession}).  This condition is much less demanding for
	concave surfaces, such as those showing high albedo in Arrokoth,
	than for convex surfaces.  {Consistent with the findings of
	\citet{L20}, the more volatile CO, N$_2$ and CH$_4$ are unlikely} to
	be retained to the present day (see discussion in text), {while
	the less volatile} CH$_3$OH, HCN and H$_2$O are unlikely to provide
	enough sublimation and redeposition to make the observed regions of
	high albedo.  NH$_3$ and C$_3$H$_8$ could sublime and condense as
	pure ices on shaded surfaces; NH$_3$ is likely to be more abundant.
	Triple points, where the inclusion of the latent heat of melting
	produces small discontinuities in slope, are indicated by dots; when
	not shown they are outside the range of the plot.  { These results
	are consistent with those of \citet{L20}.}}
\end{figure}

Even at the low temperatures of Arrokoth, the volatiles He, H$_2$, CO,
N$_2$, O$_2$, CH$_4$, C$_2$H$_6$ and possibly CO$_2$ are rapidly lost.
Arrokoth's escape velocity $v_{esc} \approx 5\,$m/s if it has the nominal
density of 0.5 g/cm$^3$ \citep{S20}, far below the thermal velocity of any
light molecule ($v_{th} \approx 150\text{--}200\,$m/s, depending on
molecular weight, at 60 K), so gravity does not much slow their escape 
{\citep{CZ09,CZ20}}.  The vapor pressures of molecules of possible interest
are shown in Fig.~\ref{Pvap}.

Vapor pressures are calculated from the triple point pressures and the
Clausius-Clapeyron equation with one component a perfect gas:
\begin{equation}
	\label{CC}
	{d\ln{P_{vap}} \over dT} = {L(T) \over k_B T^2}, 
\end{equation}
where $L(T)$ is the latent heat of the phase transition.  Not all the
required $L(T)$ are available (in particular, for NH$_3$ hydrate, that we do
not consider explicitly).  {The thermodynamic data have been reviewed
and fitted by \citet{FS09}; for NH$_3$ from \citet{OVD18,K24,OG37}.  They
have been applied to the long-term stability of Arrokoth's ices by
\citet{L20}.}

The latent heat of sublimation of CO$_2$ is obtained from \citet{GE37}. 
The latent heat of sublimation of a solid at its triple point is the sum
of the latent heat of melting and that of evaporation of the liquid; hence
$L(T)$ is discontinuous at a triple point.  This produces small
discontinuities in the slopes of the curves in Fig.~\ref{Pvap}.

When the latent heat is not directly measured (such as at temperatures below
the triple point where the vapor pressure is small), it is obtained from the
latent heat at a temperature where it is known, typically the triple point,
using
\begin{equation}
	\label{dLdT}
	{dL(T) \over dT} = C_{p,g} - C_{p,s}.
\end{equation}
The specific heat of NH$_3$ vapor, $C_{p,g} \approx 4 k_B$ per molecule, is
that of a classical gas with three rotational but no vibrational degrees of
freedom (the rotation constants are, in temperature units, 14.3 K and 8.9 K
\citep{BP57}, sufficiently below the temperatures of interest).  The
inversion level spacing of 1.14 K \citep{PMB71} contributes negligibly to
the specific heat.  The binding energy of the ammonia dimer \citep{LP00} is
about half the latent heat of evaporation of a single molecule; the Saha
equation indicates that the fraction of dimers in the saturated vapor at 60
K is about $10^{-7}$, so their effect is negligible.  At 40 K, the typical
temperature of interest, $C_{p,s} \approx 5 \times 10^6$ erg/g-K
\citep{PMB71}; for H$_2$O ice $C_{p,s} \approx 3 \times 10^6$ erg/g-K
\citep{GS36}.

The corrections to $L(T)$ of NH$_3$ contributed by Eq.~\ref{dLdT} amount to
only a few percent, and change sign around $T = 129\,$K so their effects on
the calculated vapor pressure at $\sim 60$ K approximately cancel.  This
small correction is not made for other molecules (for most of them
correction would not be possible because the specific heats of their solids
are not available).

\section{Sublimation Rate}
A surface (of a pure material) subliming into vacuum loses material at a
rate (molecules per unit area per unit time)
\begin{equation}
	{\dot N} = \alpha n_{vap} v_{th},
\end{equation}
where $n_{vap}$ is the equilibrium vapor number density, $v_{th}$ is its
thermal velocity and $\alpha$ is, by detailed balance, the sticking
probability of a vapor molecule striking the surface.  We adopt $\alpha =
1$, { usually a good approximation for surfaces {colder than
$0.5 L/k_B$},} in the absence of
empirical information.  {Because of the extreme temperature sensitivity
of vapor pressures in the relevant temperature ranges, the conclusions are
very insensitive to the value of $\alpha$ assumed.}  The surface recedes at
a speed
\begin{equation}
	\label{recession}
	v_{recession} = {{\dot N} \over n_{solid}},
\end{equation}
where $n_{solid} \approx 3 \times 10^{22}\,$cm$^{-3}$ for the simple
molecular solids { considered here}.

The equibrium vapor number density
\begin{equation}
	n_{vap} = {P_{vap} \over k_B T},
\end{equation}
where $P_{vap}$ is the equilibrium vapor pressure.  Combining these results,
the recession rate
\begin{equation}
	v_{recession} = {P_{vap} v_{th} \over n_{solid} k_B T} \sim
	10^{-4} {P_{vap} \over \text{dyne/cm$^2$}} \text{cm/s}.
\end{equation}

If 
\begin{equation}
	v_{recession} > v_{max} \sim {r_{Arrokoth} \over t_\odot} \sim
	10^{-11}\ \text{cm/s},
\end{equation}
where $t_\odot = 1.4 \times 10^{17}\,$s is the age of the Solar System as
well as that of the Sun, an object the size of Arrokoth is destroyed in
{less than} $t_\odot$.  This corresponds to maximum vapor pressure at 60 K
of $P_{max} \sim 10^{-7} \,$dyne/cm$^2$ (scaling $\propto T^{1/2}$).

The vapor pressure of CO$_2$ is probably too high for it to be retained for
the age $t_\odot$ of the Solar System because a significant fraction of the
surface of Arrokoth may be at temperatures above 50 K (at 55 K a 10 km
thickness would evaporate in $0.01 t_\odot$).  It is also evident that the
vapor pressures of H$_2$O and CH$_3$OH, {the latter spectrally identified
on Arrokoth \citep{G20},} are too low, {as shown by the fact that they are
far below either value of $P_{min}$ at any temperature $\le T_{max}$}, to
provide sufficient evaporation to deposit a high albedo pure ice.
\section{Deposition}
Fig.~\ref{Pvap} shows the extremely {nonlinear} sensitivity of vapor
pressures to temperature at the temperatures of interest.  As a result,
evaporation rates are almost entirely determined by the {\it highest\/}
temperatures a surface encounters.  Volatiles condense on the {cold-trap}
surfaces with the lowest maximum temperatures, those in the narrow cleft
between the two lobes of Arrokoth, its bright necklace.  Any surface exposed
to more insolation has a higher peak temperature and sublimation rate.
Transfer between these surfaces and the cleft is inefficient in the high
vacuum of the present Solar System because most evaporated molecules move
faster than escape velocity, but was much more efficient in the proto-Solar
System in which the gas density \citep{D07} was sufficient to make the
motion of sublimed molecules diffusive (Sec.~\ref{sgas}).

A minimum recession rate $v_{min}$ is required for sufficient material to be
sublimed to coat the cooler shaded parts of the body, such as the neck
between the lobes and the insides of depressions, and to increase their
albedo in the age of the Solar System $t_{SS}$.  This layer of deposited
pure ice crystals resembling hoarfrost has a high
visible albedo if its thickness is $\ge 0.3 \lambda$, where $\lambda$ is the
wavelength of Solar radiation \citep{LL12}.

The required minimum recession rate
\begin{equation}
	\label{vrecess}
	v_{recess} > v_{min} \sim {0.3 \lambda \over t_{SS} \eta} \sim
	10^{-22} \eta^{-1}\ \text{cm/s},
\end{equation}
where $\eta$ is the fraction of evaporated material that is deposited
on colder parts of the surface.  At low surface temperatures, resublimation
may be neglected because of the steepness of the vapor pressure curves.  Two
cases require consideration:

\begin{figure}
	\centering
	\includegraphics[width=3in]{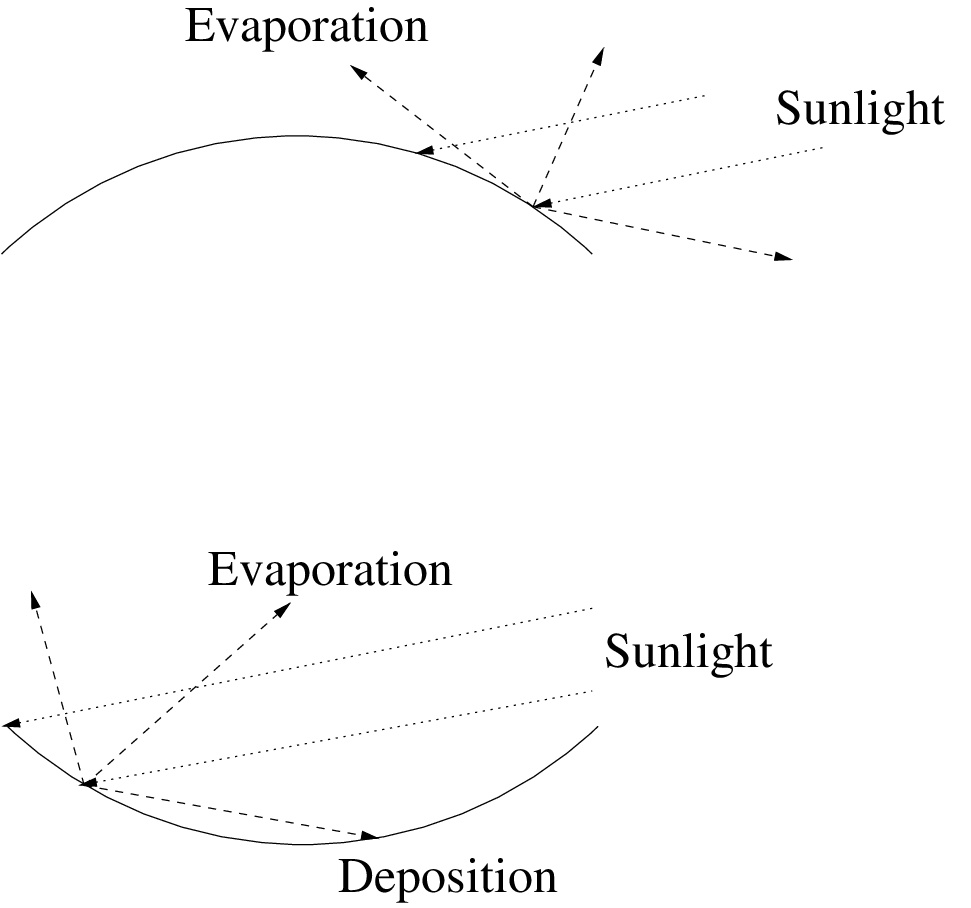}
	\caption{\label{shape} Thermal sublimation from Solar heated convex
	(above) and concave (below) surfaces in vacuum.  Dotted arrows
	indicate direction of incident sunlight, and dashed arrows the paths
	of evaporated molecules.  Molecules sublimed from concave surfaces,
	such as the cleft between the lobes of Arrokoth and in depressions
	or pits may be redeposited, even if the molecules are moving faster
	than escape speed, as almost all of them do.  However, molecules
	sublimed from convex surfaces faster than escape speed cannot
	redeposit unless the surrounding gas density is sufficient that
	their motion is diffusive, as it was in the proto-Solar System.
	Even in that case, diffusion will produce net transfer from warmer
	convexities and asperities to cooler concavities, smoothing the
	surface.}
\end{figure}
\subsection{Near-Vacuum}
{ We first consider that limit in which the gas density around Arrokoth
is so low that evaporated molecules move ballistically, almost never
colliding with another molecule.}
\subsubsection{Convex Surfaces}
\label{convex}
If the surface is convex, only sublimed molecules that have less than
escape velocity can return to colder parts of the surface and be deposited,
as shown in the upper part of Fig.~\ref{shape}.  Because $v_{esc} \ll
v_{th}$, {\citet{CZ09,CZ20}}{ only a small} fraction $\eta \sim
(v_{esc}/v_{th})^3 \sim 3 \times 10^{-5}$ (NH$_3$) or $\sim 1 \times
10^{-4}$ (CO$_2$) of the sublimed molecules remains gravitationally bound.
Only these low velocity molecules, moving ballistically, will strike the
surface again.  For NH$_3$
\begin{equation}
	\label{pvapconvex}
	P_{vap} > {0.3 \lambda n_{solid} k_B T \over t_{SS} \eta v_{th}}
	\sim 5 \times 10^{-14}\ \text{dyne/cm}^2;
\end{equation}
for CO$_2$ the required $P_{vap}$ is about 2.5 times less.
\subsubsection{Concave Surfaces}
\label{concave}
The cleft between the two lobes of Arrokoth is concave, as are the insides
of depressions.  Molecules sublimed from the warmer parts of the surface may
strike colder parts, even though their velocities exceed $v_{esc}$.  This is
illustrated in the lower part of Fig.~\ref{shape}.  Then $\eta$ may be much
higher, perhaps ${\cal O}(\text{0.1--1})$, and
\begin{equation}
	\label{pvapconcave}
	P_{vap} > 10^{-17}\ \text{dyne/cm}^2.
\end{equation}
\subsubsection{Net Deposition}
\label{net}
Matter on both the warmer and the cooler parts of the surface will sublime,
but the cooler parts will be net gainers of ice if their sublimation rate is
$< \eta$ of the evaporation rate of the warmer parts.  This condition will
be satisfied if $\Delta T > \Delta T_{min}$, where
\begin{equation}
	\label{DTmin}
	\Delta T_{min} \approx - {\ln{\eta} \over d\ln{P_{vap}}/dT};
\end{equation}
for NH$_3$ at 50 K $d\ln{P_{vap}/dT} = {L / k_B T^2} \approx \text{1.5/K}$.

For a convex surface in vacuum $\Delta T_{min} \approx 7\,$K.  The
temperature difference between unshadowed and shadowed parts of the surface
likely exceeds this because the mean temperature is 42 K, about {16--19}
K below the sub-Solar temperature { (Sec.~\ref{steadystate})}.  For a
concave surface in vacuum like the cleft $\Delta T = {\cal O}$(1 K) is
sufficient for net deposition on its colder parts.
\subsection{Immersed in Gas}
\label{sgas}
As discussed in Sec.~\ref{rotation} and by \citet{Ly20,McK20}, the
rotational state of Arrokoth suggests it was immersed in the gas of the
proto-Solar System with a density $\rho_g \sim 4 \times 10^{-11} (2 \times
10^6\, \text{y}/t_{slow})$, where $t_{slow}$ was the time scale on which gas
drag slowed its rotation.  {This process has also been inferred in
proto-planetary discs by \citet{A16,O17,Lo20} and in models of the
proto-Solar System by \citet{LD11,Kr18,Kr20,M19}.}  The collisional mean
free path of a sublimed molecule
\begin{equation}
	\label{ell}
	\ell \sim {m_g \over \rho_g \sigma} \sim 25 \left({t_{slow} \over
	2 \times 10^6\,\text{y}}\right)\,\text{cm},
\end{equation}
where $\sigma \sim 3 \times 10^{-15}$ cm$^2$ is a representative molecular
collision cross-section and the gas is assumed to have been of Solar
composition.  This length was orders of magnitude less than the size of
Arrokoth if $t_{slow}$ was in the plausible range $10^6$--$10^8$ y
\citep{St20}.  As a result, sublimed molecules did not escape ballistically
but rather diffused in the surrounding gas.  Colder parts of the surface
were effective sinks for {diffusing} sublimed molecules and the redeposition
fraction $\eta$ may have approached unity.  Then (Eq.~\ref{DTmin}), the
temperature difference required for net deposition was small.

This process may also have contributed to the smoothing of the surface of
Arrokoth by transporting matter from warmer sunlit convexities to cooler
shadowed concavities.  Because the { maximal possible} thermal relaxation
time { (assuming the highest possible bulk conductivity)} $t_{relax}$
(Eq.~\ref{relax}) is a substantial fraction of, but less than, $P_{orb}$,
the highest temperature achieved by a point on the surface (determining the
sublimation rate) depends in a continuous manner on the shape, with convex
asperities preferentially eroded (in both the ballistic and diffusive
regimes).  The fact that the cleft between the lobes was not filled
indicates that { any thermally driven deposition} process { slowed}
before the two predecessor bodies came into contact, { although the
amount of deposition required to make a bright necklace is many orders of
magnitude {($\sim 2000\,\text{\AA/2\,km} \sim 10^{-10}$)} less than that
required to reshape the body}.  

A region on which a layer $\gtrsim 0.3\lambda$ thick of ice has been
deposited has a much higher visible albedo than a dirty ice surface.  The
albedo of ice or snow depends on its structure (a continuous solid or many
small crystals separated by voids).  Analogy with terrestrial water-ice
hoarfrost suggests a low density porous structure with an albedo that may
approach unity because of the absence of absorbing mineral matter.  Its
temperature drops (the magnitude of the decrease depends not only on its
albedo but on its proximity to low albedo dirty ice regions, from which heat
conducts), accelerating the net deposition of ice.

A surface containing transparent volatiles is subject to an instability in
which the volatiles evaporate from warmer, lower albedo, regions and deposit
on cooler, higher albedo, regions, further increasing their albedo.  Patches
of high albedo ``hoarfrost'' grow at the expense of darker regions.  
Because of the extreme sensitivity of vapor pressure to temperature, these
``hoarfrost'' regions will be chemically pure, consisting of only the one
chemical species that does not evaporate where it is deposited in the cooler
shaded regions while being volatile enough to evaporate from warmer regions.

{On a chemically pure surface Solar ultraviolet and cosmic ray
irradiation cannot make the darker ``tholins'' that result from their action
on a mixture of carbonaceous and nitrogenous species, forming C--N bonds.
The same shadowing mechanism that keeps these regions comparatively cool
also reduces the rate at which their material is reprocessed into darker
compounds, and may preclude it entirely.  Darkening by photochemical
reprocessing competes with lightening by vapor deposition of ``snows''.
Shadowing tilts this competition against darkening, both by reduction of
irradiation and by providing a pure, single-species, composition.}
\section{Rotation}
\label{rotation}
The rotation period of Arrokoth is 15.92 hours \citep{S19}.  The
orbital period of two homogeneous spheres of density $\rho$ in contact, an
approximate description of its (imperfectly known) shape, is
\begin{equation}
	P_{breakup} = \sqrt{{3 \pi \over G\rho}{(1+\zeta)^3 \over
	1+\zeta^3}} \approx 9.2\,\text{h},
\end{equation}
where $\zeta = 0.81$ \citep{S20} is the ratio of the spherical equivalent
radius of the smaller sphere to that of the larger sphere and a nominal
density $\rho = 0.5$ g/cm$^3$, { as found for comets and for other small
Kuiper Belt Objects \citep{McK20,S20},} is assumed.  If the lobes
are not spherical, the mean density of spheres in contact whose centers of
masses coincide with those of the lobes should be substituted for $\rho$
(the result would be a good approximation, but not exact).  $P_{breakup}$ is
the shortest possible rotation period of a strengthless contact binary, and
is about 0.6 of the actual rotation period, implying that the contact is in
compression.  Possible { dominant} constituents have { bulk} densities
at low temperature (NH$_3$ 0.82 g/cm$^3$ \cite{B75}, H$_2$O 0.92 g/cm$^3$,
CO$_2$ 1.6 g/cm$^3$, { CH$_3$OH 1.02 g/cm$^3$ \citep{D56}, HCN 1.04
g/cm$^3$ \citep{R69}}) higher than the assumed nominal density $\rho_{Ar}$
of Arrokoth; the nominal density, observed for comets, implies a structure
{ with high porosity} \citep{McK20,S20}.

The shape of Arrokoth does not conform to the equipotential (Roche) surfaces
of a uniformly rotating fluid with mass concentrated at two orbiting points,
in which each lobe is drawn to a singularity at their contact.  Nor does it
resemble the highly prolate (Jacobi ellipsoid) shape of homogeneous
self-gravitating bodies of high angular momentum in equilibrium, such as
inferred for `Oumuamua \citep{Me17,K18} {and Haumea \citep{D19}}.  Nor
does it resemble the oblate McLauren spheroids of lower angular momentum,
{as suggested for Quaoar \citep{BR13}}.  This argues against formation
models in which a single object was formed by angular momentum-limited
accretion (as in an accretion disc like those observed in mass-transfer
binary stars) from surrounding gas or particles.  The problem of explaining
the present state of Arrokoth is the opposite of that of explaining contact
and close binary asteroids that are apparently spun up by the YORP effect
(that is small in the Kuiper Belt) or by planetary encounters that do not
occur for cold classical Kuiper Belt objects in near-circular orbits
\citep{S07}.

This suggests that the two lobes formed independently as slowly rotating
(slower than their breakup rotation rate) objects.  Orbiting but separated,
their orbit gradually contracted as it lost angular momentum until they came
into contact.  The long rotation period suggests that some process continued
to remove angular momentum after Arrokoth formed, plausibly the same
process that also brought its lobes together by removing angular momentum.
Even in vacuum, Solar gravity induces Lidov-Kozai oscillations that may
bring the lobes into contact \citep{Gr20}; we only consider processes that
are related to the transfer of vapor from one part of the body to another
that may explain the bright necklace.

Angular momentum could have been removed by hydrodynamic interaction with
surrounding gas in the proto-Solar System \citep{D07,Ly20,McK20}.
Interaction with gas of density $\rho_g$ produces a torque $\sim \rho_g v^2
r^3 \sim G M_{Ar} \rho_g r^2$, where $r$ is the radius of Arrokoth, its
mass is $M_{Ar}$  and the speed of its lobes through the gas $v \sim
\sqrt{GM_{Ar}/r}$.  Gravitational interaction with the gas produces a
comparable characteristic torque $\sim (GM_{Ar}/r^2) M_g r \sim G M_{Ar}
\rho_g r^2$.  The slowing time
\begin{equation}
	t_{slow} \sim {1 \over \rho_g} \sqrt{\rho_{Ar} \over G} \sim 2
	\times 10^6 \left({\rho_g \over 4 \times 10^{-11} \text{g/cm}^3}
	\right),
%	\rho_g \sim {1 \over t_{slow}}\sqrt{\rho_{UT} \over G} \approx
%	{3 \times 10^3\,\text{g-s/cm}^3 \over \rho_g} \approx 10^{-11}\,
%	\text{g/cm}^3 {10^7\,\text{y} \over t_{slow}}.
\end{equation}
where the gas density has been scaled to that of 1 $M_\odot$ of gas
uniformly filling an oblate spheroid of major radii $a = 44$ AU and axial
ratio equal to the $2.45^\circ = 0.043$ radian inclination of Arrokoth.  The
end state of removal of angular momentum by gas drag is consistent with the
observed contact binary.  {This gas is almost entirely hypervolatile
hydrogen and helium, and does not condense.  Any carbon, nitrogen and
oxygen, if not yet reduced to hydrocarbons, ammonia, and water, might
condense as CO, N$_2$ or O$_2$ but these volatile species would quickly
evaporate afterward as the proto-Solar System became transparent to Solar
radiation \citep{GW18,McK20,L20,St20}.  In our hypothesis, the bright
necklace is produced later.}

As the gaseous environment removes angular momentum, bringing the
predecessor bodies into contact and then slowing the rotation below the
breakup rate \citep{McK20}, so also the proto-Arrokoth exerted a torque on
the gas, driving it away, in analogy to the satellite-driven gaps in
planetary rings.  The end state, a close binary or bilobate single object
like Arrokoth, depends on the resupply of gas and therefore on the poorly
understood processes of angular momentum transport in the proto-Solar System
gas cloud.
\section{Discussion}
\label{discussion}
The values of $P_{vap}$ corresponding to recession rates $v_{min}$ and
$v_{max}$ are indicated in Fig.~\ref{Pvap}.  NH$_3$ and C$_3$H$_8$
have vapor pressures that would permit deposition of high albedo ice in
shaded regions of Arrokoth but low enough that they would not be lost
entirely during the age of the Solar System; CO$_2$ is a marginal member of
this class.  Of these, NH$_3$ is expected to be the most abundant and to be
a significant constituent of Kuiper Belt objects \citep{B12}.  {These
finely divided ices (snows) have high albedos across the visible spectrum
\citep{P92,H21}, consistent with the high albedo of Arrokoth's necklace.
The absence of NH$_3$ in the infrared reflection spectrum of Arrokoth
{outside the necklace} \citep{G20} may be explained by photochemical
processing to ``tholins'', also explaining its red color, but the spatial
resolution of these spectra was insufficient to constrain the spectrum of
{any ice in the narrow necklace.}}

Because of the steep dependence of vapor pressure on temperature at 
$k_B T \sim L(T)$ (Fig.~\ref{Pvap}; this is true of any substance when its
vapor pressure is small { because of the self-similar scaling of the
Clausius-Clapeyron equation}), sublimation chiefly occurs where the peak
insolation is greatest and the maximum temperatures are the highest.  Net
condensation chiefly occurs where the maximum temperature is the lowest;
minimum temperatures make little difference because even at temperatures
only slightly less than the highest a subliming region experiences the
sublimation rate is negligible.  The cleft between the lobes of Arrokoth is
sunlit only for brief {intervals, of width $\sim P_{orb} \theta/2\pi$
that may be $\sim 0.03 P_{orb} \sim 10\,$y} around the equinoxes.  Because
$t_{relax}$ is comparatively long, the temperature in the cleft remains
closer to the deep interior temperature $\langle T \rangle$ than it does
anywhere else on the surface.  As a result, the cleft traps volatiles
sublimed elsewhere as pure ices, giving it a high albedo (and hence yet
lower surface temperature), and the bright ``necklace'' evident in the
encounter images \citep{S19}.

The fact that Arrokoth consists of two lobes in contact, in contrast to
apparently highly prolate 1I/2017 U1 `Oumuamua \citep{Me17,B18,O19}, 
{suggests} a different origin and history.  `Oumuamua may have acquired its
shape and been expelled from its parent planetary system during a luminous
post-main sequence phase of its star \citep{K18}, {or it could have been
tidally elongated during a close, ejecting passage by a forming exo-Jovian
planet \citep{B18}.  In {analogy to the latter possibility}, Arrokoth
may have formed and acquired its present configuration within the Kuiper
belt when the proto-Solar System was filled with comparatively dense gas
{within the first $\sim 10^6\text{--}10^7$ y of the Solar System's
existence \citep{R05,M09,R18}}.}

The process that smoothed the lobes of Arrokoth must have ceased before they
came into contact, because Arrokoth itself retained the deep cleft between
its lobes.  If $P_{vap} \gg P_{max}$ before the lobes came into contact,
then diffusive transport in the comparatively dense proto-Solar System
nebula could have effectively smoothed them in much less than the age of the
Solar System.  The higher temperatures of more insolated convexities and the
lower temperatures of more shadowed concavities would round the lobes as
matter would preferentially be lost from convexities.  This process would
have been particularly effective in eroding asperities, that are more open
to insolation and less closely conductively coupled to the interior.

In the comparatively dense gas of the proto-Solar System \citep{D07} {
($n \gtrsim 10^9$ cm$^{-3}$, corresponding to a mass $\sim 10^{30}$ g, is
sufficient to give mean free paths $\lesssim 3$ km; Sec.~\ref{sgas})} so
that sublimed molecules do not escape on ballistic trajectories.  Instead,
they diffuse in the surrounding gas and are readily recondensed on colder
surfaces, with $\eta$ of order unity.  { The value of $\eta$ may be
comparable to that estimated in Sec.~\ref{concave}.  To produce a high
albedo necklace during this phase would require $P_{vap}$ greater than that
of Eq.~\ref{pvapconcave} by a factor of $10^3\text{--}10^4$, the ratio of
the age of the Solar System (used in Eq.~\ref{pvapconcave}) to the duration
of its stage as a dense proto-planetary disc, and numerically similar to the
value (Eq.~\ref{pvapconvex}) for a convex surface.}
%For this to be effective for NH$_3$
%in $\sim 10^6\text{--}10^7\,$y would require $T \approx \text{80--90}$ K,
%consistent with a proto-Sun on its Hayashi track
%\citep{KWW12}.  As the proto-Sun
%settled onto the zero-age Main Sequence, its decreasing luminosity would
%have led to a reduction in the surface temperature of Arrokoth and a great
%reduction in the sublimation rate, unless the temperature had previously
%been reduced by opacity in the orbital plane \citep{G20}.
The dissipation of the proto-Solar System gas cloud meant that sublimed
molecules would (except in concavities) almost all escape ballistically
rather than redepositing.

{These arguments suggest that Arrokoth's shape, slow spin, and geometry
of its necklace formed in the early proto-Solar System.  The observed highly
reflective and perhaps very thin (sub-micron) hoarfrost necklace may be much
younger ($\lesssim 10^8\,$y).  Close passage of luminous stars or nearby
supernova\ae\ \citep{S03,L20} may evaporate an existing thin reflective
layer that would be subsequently redeposited.  Such shorter ages would imply
that the relevant pressures are higher than shown in Fig.~\ref{Pvap}, but
the dependence of vapor pressure on temperature is so steep
($d\ln{P_{vap}}/dT \sim 1/$K) that the changes in the corresponding
temperatures would be small.  The gradual increase in the Solar luminosity
with age produces a similar effect: the vapor pressure of NH$_3$ at
$T_{max}$ has an $e$-folding time $\sim 3 \times 10^8\,$y, a time scale that
may be more relevant than the age of the Solar System.} 
\section*{Appendix}
The equilibrium temperature of a Solar System object is determined by the
balance between Solar heating and its infrared thermal emission.  This
depends on its albedo for Solar radiation and its infrared emissivity
averaged over a Planck function at its surface temperature.  Small bodies in
the Kuiper Belt and elsewhere in the outer Solar System, such as Arrokoth
(486958; 2014 MU$_{69}$) and elsewhere in the outer Solar System { are
made of rock and ices}.  Vapor pressures are extremely sensitive to
temperature at the temperatures $\approx$~50~K found there, so quantitative
determination of temperature is necessary to calculate the rate of
evaporative loss and vapor transport across their surfaces.  This requires
quantitative knowledge of the infrared emissivity.

Ammonia ice is of particular interest because in this temperature range its
vapor pressure varies from values so low that it is insufficient to deposit
sub-micron thick layers on colder parts of a body's surface in the age of
the Solar System to values so large that a 10 km-sized body would not
survive that time.  In contrast, the vapor pressures of water
and methanol ices are so low that evaporation is negligible, while those of
methane and even carbon dioxide are so high that they would be lost entirely
from a body with negligible gravity.

We use the complex infrared optical constants of ammonia ice \citep{MOA84}
to calculate its reflectivity $R(\nu,\theta,{\hat n})$ as a function of
frequency $\nu$, angle of incidence $\theta$ and polarization $\hat
\epsilon$ from the Fresnel relations.  The emissivity
$\epsilon(\nu,\theta,{\hat \epsilon}) = 1 - R(\nu,\theta,{\hat \epsilon})$.

We assume a homogeneous half-space of solid ammonia; transmitted energy is
eventually absorbed, either by the imaginary part of the refractive index
(which is very small in most of the spectrum) or by embedded mineral matter
(dirt).  If this assumption were not made it would be necessary to specify
the depth of the ice layer, which is not known, and to match electromagnetic
boundary conditions at both interfaces.  In fact, the low ($\approx 0.1$)
visible albedo of Arrokoth \citep{S19} implies a deep homogeneous layer
of ice in which some mineral matter is embedded; if the ice were finely
divided, like snow, the visible albedo would be high because of scattering
at interfaces between ice and vacuum (or air, for terrestrial snow).

Once the mean infrared reflectivity $\langle R(T) \rangle$, averaged over a
Planck function, is known, the equilibrium temperature may be calculated.
There are two simple cases.  If sunlight is normally incident with intensity
$I_\odot$ the steady state temperature is
\begin{equation}
        \label{normal}
        T_{normal} = \left({I_\odot (1-A) \over \sigma_{SB} (1-\langle R
        \rangle)}\right)^{1/4},
\end{equation}
where $A$ is the Solar albedo, averaged over its spectrum, and $\sigma_{SB}$
is the Stefan-Boltzmann constant.  Averaging over a spherical body yields a
mean temperature
\begin{equation}
        \label{mean}
        T_{mean} = \left({I_\odot (1-A) \over 4 \sigma_{SB} (1-\langle R
        \rangle)}\right)^{1/4},
\end{equation}

The Fresnel relations for interfaces between dielectric (non-magnetic)
materials are
\begin{equation}
        R_s = \left\vert{\sqrt{1-\sin^2\theta} - \sqrt{n^2-\sin^2\theta}
        \over \sqrt{1-\sin^2\theta} + \sqrt{n^2-\sin^2\theta}}\right\vert^2
\end{equation}
and
\begin{equation}
        R_p = \left\vert{n^2\sqrt{1-\sin^2\theta} - \sqrt{n^2-\sin^2\theta}
        \over n^2\sqrt{1-\sin^2\theta} + \sqrt{n^2-\sin^2\theta}}\right\vert^2,
\end{equation}
where $s$ denotes polarization in the plane of incidence, $p$ denotes
polarization perpendicular to the plane of incidence and $n(\nu)$ is the
complex relative refractive index (the index of the solid when the wave is
incident from vacuum).

\begin{figure}
        \centering
        \includegraphics[width=\columnwidth]{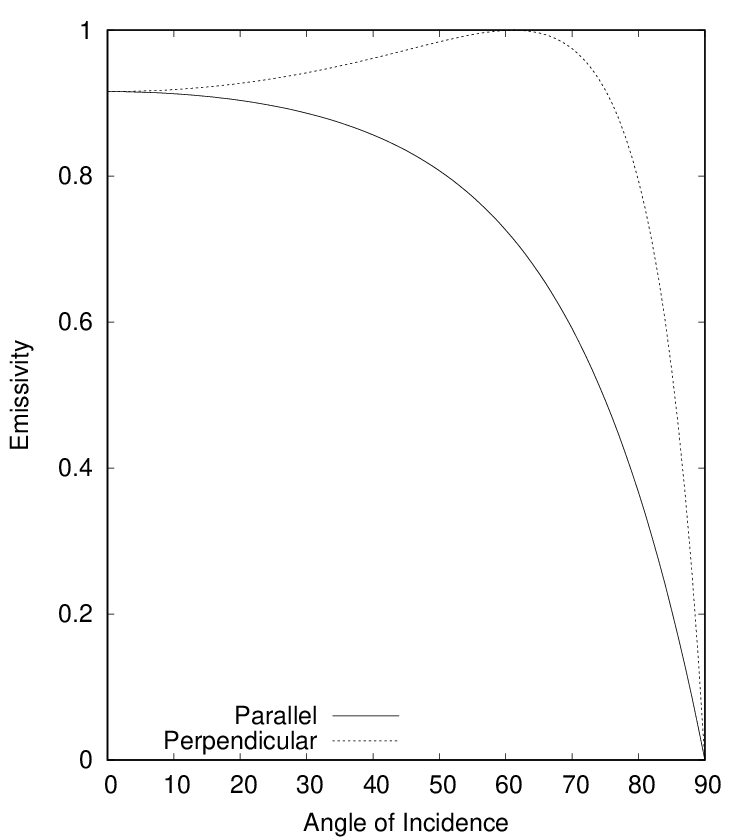}
        \caption{\label{theta} Emissivity ($1-R$) for infrared radiation
        with $\lambda = 50\mu$ as a function of angle of incidence.}
\end{figure}

\begin{figure}
        \centering
        \includegraphics[width=\columnwidth]{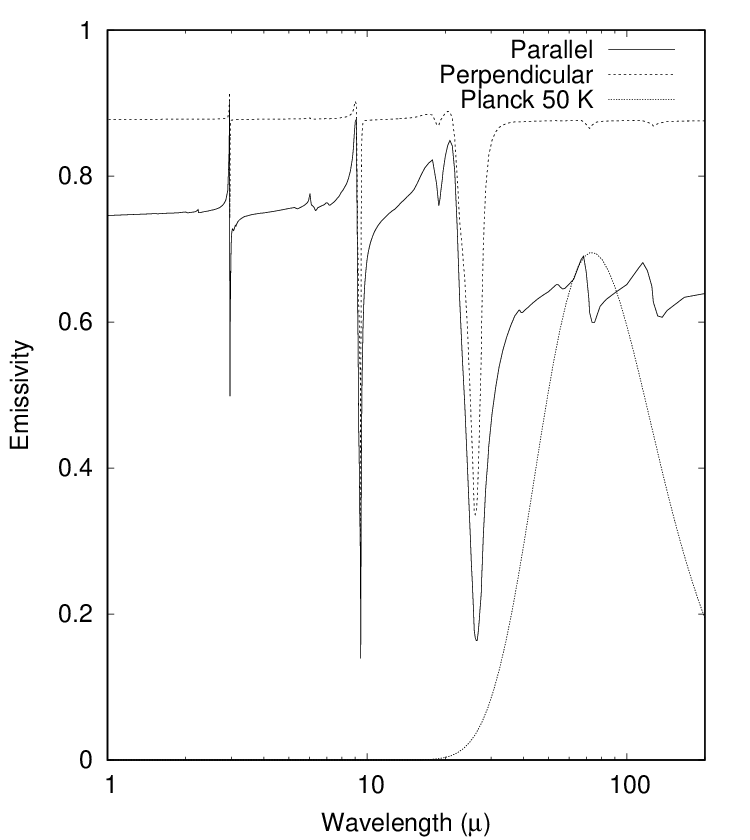}
        \caption{\label{nu} Emissivity ($1-R$) as a function of wavelength
        and polarization, averaged over solid angles.  The Planck function
        $F_\lambda$ at 50 K, with arbitrary normalization, is shown for
        comparison.}
\end{figure}

\begin{figure}
        \centering
        \includegraphics[width=\columnwidth]{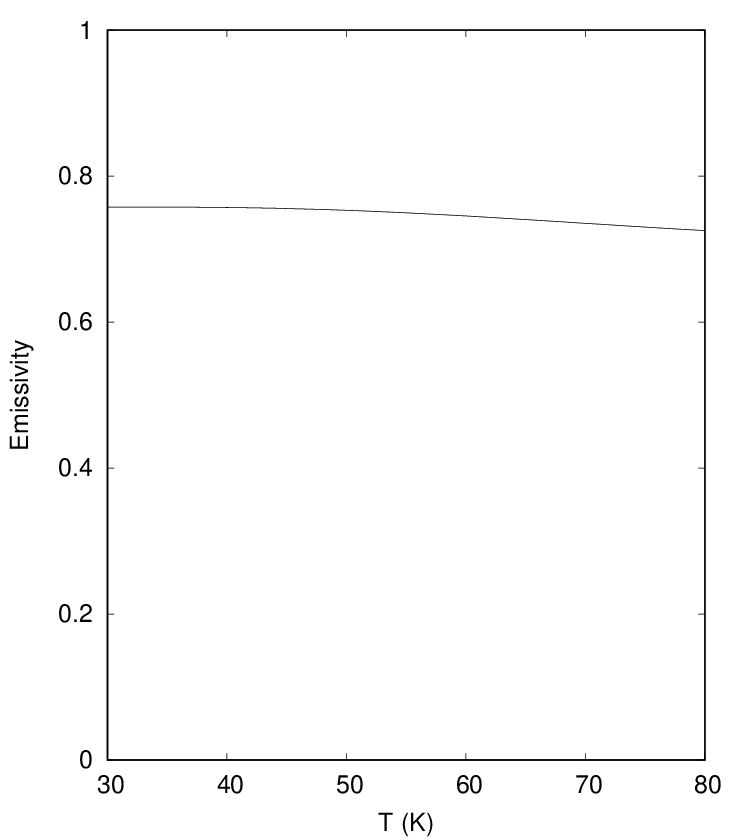}
        \caption{\label{T} Frequency-integrated emissivity ($1-\langle R
        \rangle$) as a function of temperature, averaged over solid angles,
        polarization and Planck function.}
\end{figure}

The mean reflectivity
\begin{equation}
        \langle R(T) \rangle = {1 \over 2} {\int_0^\infty\!d\nu\int_1^0\!
        d\cos\theta [R_s(\nu,\theta) + R_p(\nu,\theta)] F_\nu(T) \over
        \int_0^\infty\!d\nu\int_1^0\!d\cos\theta F_\nu(T)},
\end{equation}
where $F_\nu$ is the Planck function.  While $R_s$ and $R_p$ are properties
of the material, $R(T)$ depends on its temperature through $F_\nu(T)$.  The
results are shown in Figures \ref{theta}--\ref{T}.

The Planck-averaged emissivity at temperatures of interest for the Kuiper
Belt, where solid ammonia is likely to be encountered, is $\approx 0.77$ and
varies only slightly with temperature.  Inclusion of this factor in
Eqs.~\ref{normal} and \ref{mean} yields a temperature about 7.5\% higher
than would be estimated if the body were a black body radiator.  The vapor
pressure is so sensitive to temperature that this can be significant.

Finely divided ammonia ``snow'' has higher reflectivity (lower emissivity)
because light scatters at every interface between solid and vacuum.  This is
analogous to the high albedo of terrestrial water-snow.  However, it is less
extreme, because at wavelengths $\lambda \sim 100\mu$ near the 50 K black
body peak the imaginary part of the refractive index $n_i \sim 0.1$ but
varies rapidly with $\lambda$ \citep{MOA84}, suggesting $1-R \sim 0.1$.
Multiple scatterings further reduce $R$, but a quantitative calculation
would require detailed knowledge of the geometry.  However, this factor
is offset by the fact that the Solar albedo $A$ of pure ammonia snow is
likely to be close to unity (\citet{MOA84} give $n_i = 2\text{--}4 \times
10^{-5}$ for blue and red light).  This is in contrast to the Solar albedo
of bulk solid ammonia, which is likely to be small because of mineral or
carbonaceous contamination (``dirt''), consistent with the low albedo of
Arrokoth.  Qualitatively, pure ammonia snow will be significantly cooler
than solid ammonia, so that if the the surface is warm enough to evaporate
vapor, vapor-deposited ice will accumulate further material at
the expense of solid ammonia.

\section*{Acknowledgement}
We thank L. M. Canel-Katz, D. Eardley, R. Ogliore, J. R. Spencer and
anonymous referees for useful discussions.

\section*{Data Availability}
This theoretical paper does not contain any new data.

\bsp %MNRAS
\label{lastpage}

\begin{thebibliography}{99}
	\bibitem[\protect\citeauthoryear{Anderl {\it et al.\/}}{2016}]{A16}
		Anderl, S., Manet, S., Cabrit, S. {\it et al.\/} 2016 \aap\
		591A, 3.
	\bibitem[\protect\citeauthoryear{Benedict \& Plyler}{1957}]{BP57}
		Benedict, W.~S. \& Plyler, E.~K. 1957 Can.~J.~Phys. 35,
		1235.
	\bibitem[\protect\citeauthoryear{Blum}{1975}]{B75} Blum, A. 1975
		Radiation Effects and Defects in Solids 24, 277.
	\bibitem[\protect\citeauthoryear{Bolin {\it et al.\/}}{2018}]{B18}
		Bolin, B. T., Weaver, H. A., Fernandez, Y. R. {\it et al.\/}
		2018 \apjl\ 852, L2.
	\bibitem[\protect\citeauthoryear{Braga-Ribas {\it et al.\/}}{2013}]
		{BR13} Braga-Ribas, F., Sicardy, B., Oritz, J. L. {\it et
		al.\/} 2013 \apj\ 773, 26.
	\bibitem[\protect\citeauthoryear{Brown}{2012}]{B12} Brown, M. E.
		2012 Ann. Rev. Earth Planet. Sci. 40, 467.
	\bibitem[\protect\citeauthoryear{Catling \& Zahnle}{2009}]{CZ09}
		Catling, D. C. \& Zahnle, K. J. 2009 Scientific American
		300, 36.
	\bibitem[\protect\citeauthoryear{Catling \& Zahnle}{2020}]{CZ20}
		Catling, D. C. \& Zahnle, K. J. 2020 Sci. Adv. 6, eaax1420.
	\bibitem[\protect\citeauthoryear{Davidson}{1956}]{D56} Davidson, D.
		W. 1956 Can. J. Chem. 34, 1243. 
	\bibitem[\protect\citeauthoryear{Desch}{2007}]{D07} Desch, S. J.
		2007 \apj\ 671, 878.
	\bibitem[\protect\citeauthoryear{Dunham, Desch \& Probst}{2019}]
		{D19} Dunham, E. T., Desch, S. J. \& Probst, L. 2019 \apj\
		877, 41.
	\bibitem[\protect\citeauthoryear{Fray \& Schmitt}{2009}]{FS09} Fray,
		N. \& Schmitt, B. 2009 Plan.~Sp.~Sci. 57, 2053.
	\bibitem[\protect\citeauthoryear{Giauque \& Egan}{1937}]{GE37}
		Giauque, W. F. \& Egan, C. J. 1937 J.~Chem.~Phys. 5, 45.
	\bibitem[\protect\citeauthoryear{Giauque \& Stout}{1936}]{GS36}
		Giauque, W. F. \& Stout, J. W. 1936 J.~Am.~Chem.~Soc. 58,
		1144.
	\bibitem[\protect\citeauthoryear{Glein \& Waite}{2018}]{GW18} Glein,
		C. R. \& Waite, J. H. 2018 Icarus 313, 79.
	\bibitem[\protect\citeauthoryear{Grishin {\it et al.\/}}{2020}]
		{Gr20} Grishin, E., Malamud, U., Perets, H. B., Wandel, O.
		\& Sch{\"a}fer, C. M. 2020 Nature 580, 463.
	\bibitem[\protect\citeauthoryear{Grundy {\it et al.\/}}{2020}]{G20}
		Grundy, W. M., Bird, M. K., Britt, D. T. {\it et al.\/} 2020
		Science 367, eaay3705.
	\bibitem[\protect\citeauthoryear{Hofgartner {\it et al.\/}}{2021}]
		{Ho21} Hofgartner, J. D., Buratti, B. J., Benecchi, S. D.
		{\it et al.\/} 2021 Icarus 356, 113723.
	\bibitem[\protect\citeauthoryear{Hudson {\it et al.\/}}{2021}]{H21}
		Hudson, R. L., Gerakines, P. A., Yarnall, Y. Y. \& Coones,
		R. T. 2021 Icarus 354, 114033.
	\bibitem[\protect\citeauthoryear{Karwat}{1924}]{K24} Karwat, E. 1924
		Z.~Phys.~Chem. 112, 486.
	\bibitem[\protect\citeauthoryear{Kastner {\it et al.\/}}{1997}]{K97}
		Kastner, J. H., Zuckerman, B., Weintraub, D. A. \&
		Forveille, T. 1997 Science 277, 67.
	\bibitem[\protect\citeauthoryear{Katz}{2018}]{K18} Katz, J. I.
		2018 MNRAS Lett. 478, L95.
%	\bibitem[\protect\citeauthoryear{Keane {\it et al.\/}}{2020}]{K20}
%		Keane, J. T., Porter, S. B., Beyer, R. A. {\it et al.\/}
%		2020 Bull. AAS DPS 52, 508.02.
	\bibitem[\protect\citeauthoryear{Kippenhahn, Weigert \& Weiss}
		{2012}]{KWW12}Kippenhahn, R., Weigert, A. \& Weiss, A. 2012
		Stellar Structure and Evolution 2nd ed. Springer, Berlin.
	\bibitem[\protect\citeauthoryear{Korolyuk {\it et al.\/}}{2009}]
		{KKSR09} Korolyuk, O. A., Krivchikov, A. I., Sharapova, I.
		V. \& Romantsova. O. O. 2009 Fizika Nizkikh Temperatur 35,
		380.
	\bibitem[\protect\citeauthoryear{Krijt {\it et al.\/}}{2018}]{Kr18}
		Krijt, S., Schwarz, K. R., Bergin, E. A. {\it et al.\/} 2018
		\apj\ 864, 78.
	\bibitem[\protect\citeauthoryear{Krijt {\it et al.\/}}{2020}]{Kr20}
		Krijt, S., Bosman, A. D., Zhang, K. {\it et al.\/} 2020
		\apj\ 899, 134.
	\bibitem[\protect\citeauthoryear{Krupskii, Manzhely \& Koloskova}
		{1968}]{KMK68} Krupskii, I. N., Manzhely, V. G. \&
		Koloskova, L. A. 1968 Physica Status Solidi 27, 263.
	\bibitem[\protect\citeauthoryear{Lee \& Park}{2000}]{LP00} Lee,
		J.~S. \& Park, S.~Y. 2000 J. Chem. Phys. 112, 230.
	\bibitem[\protect\citeauthoryear{Lesniak \& Desch}{2011}]{LD11}
		Lesniak, M. V. \& Desch, S. J. 2011 \apj\ 740, 118.
	\bibitem[\protect\citeauthoryear{Lisse {\it et al.\/}}{2021}]{L20}
		Lisse, C. M., Young, L. A., Cruikshank, D. P. {\it et al.\/}
		2021 Icarus 356, 114072.
		\url{https://doi.org/10.1016/j.icarus.2020.114072}
		arXiv:2009.02277.
	\bibitem[\protect\citeauthoryear{Lock \& Laven}{2012}]{LL12} Lock,
		J.~A. \& Laven, P. 2012 J.~Opt.~Soc.~Am. 29, 1489.
	\bibitem[\protect\citeauthoryear{Loomis {\it et al.\/}}{2020}]{Lo20}
		Loomis, R. A., {\"O}berg, K. I., Andrews, S. M. {\it et
		al.\/} 2020 \apj\ 893, 101.
	\bibitem[\protect\citeauthoryear{Lyra, Youdin \& Johansen}{2021}]
		{Ly20} Lyra, W., Youdin, A. N. \& Johansen, A. 2021 Icarus
		356, 113831 arXiv:2003.00670.
	\bibitem[\protect\citeauthoryear{Mamajek}{2009}]{M09} Mamajek, E. E.
		2009 AIP Conf. Proc. 1158, 3.
	\bibitem[\protect\citeauthoryear{Martonchik, Orton \& Appleby}
		{1984}]{MOA84} Martonchik, J. V., Orton, G. S. \& Appleby,
		J. F. 1984 Appl. Optics 23, 541.
	\bibitem[\protect\citeauthoryear{McKinnon {\it et al.\/}}{2020}]
		{McK20} McKinnon, W. B., Richardson, D. C., Marohnic, J. C.
		{\it et al.\/} 2020 Science 367, eaay6620.
	\bibitem[\protect\citeauthoryear{Meech {\it et al.\/}}{2017}]{Me17}
		Meech, K. J., Weryk, R., Micheli, M. {\it et al.\/} 2017
		Nature 552, 378.
	\bibitem[\protect\citeauthoryear{Mousis {\it et al.\/}}{2019}]{M19}
		Mousis, O., Ronnet, T. \& Lunine, J. I. 2019 \apj\ 875, 9.
	\bibitem[\protect\citeauthoryear{Nayakshin {\it et al.\/}}{2020}]
		{N20} Nayakshin, S., Tsukagoshi, T., Hall, C. {\it et al.\/}
		2020 \mnras\ 495, 285.
	\bibitem[\protect\citeauthoryear{{\"O}berg {\it et al.\/}}{2017}]
		{O17} {\"O}berg, K. I., Guzm{\'a}n, V. V., Merchantz, C. J.
		{\it et al.\/} \apj\ 839, 43.
	\bibitem[\protect\citeauthoryear{Osborne \& Van Dusen}{1918}]{OVD18}
		Osborne, N.~S. \& Van Dusen, M.~S. 1918 Bull.~Bur.~Stand.
		14, 439.
	\bibitem[\protect\citeauthoryear{`Oumuamua ISSI Team}{2019}]{O19}
		`Oumuamua ISSI Team Bannister, M. T., Bhandare, A.,
		Dybczy\'{n}ski, P. T. {\it et al.\/} 2019 Nature Astr. 3,
		594.
	\bibitem[\protect\citeauthoryear{Overstreet \& Giauque}{1937}]{OG37}
		Overstreet, R. \& Giauque, W. F. 1937 J.~Am.~Chem.~Soc. 59,
		254.
	\bibitem[\protect\citeauthoryear{Pope {\it et al.\/}}{1992}]{P92}
		Pope, S. K., Tomasko, M. G., Williams, M. S., Perry, M. L.,
		Doose, L. R. \& Smith, P. H. 1992 Icarus 100, 203.
	\bibitem[\protect\citeauthoryear{Popov, Manzhelii \& Bagatskii}
		{1971}]{PMB71} Popov, V.~A., Manzhelii, V.~G. \& Bagatskii,
		M.~I. 1971 J.~Low Temp.~Phys. 5, 427.
	\bibitem[\protect\citeauthoryear{Porter {\it et al.\/}}{2018}]{P18}
		Porter, S. B., Buie, M. W., Parker, A. H. {\it et al.\/}
		2018 Astron. J. 156, 20.
	\bibitem[\protect\citeauthoryear{Rae}{1969}]{R69} Rae, A. I. M.
		1969 Mol. Phys. 16, 257.
	\bibitem[\protect\citeauthoryear{Richert {\it et al.\/}}{2018}]{R18}
		Richert, A. J. W., Getman, K. V., Feigelson, E. D. {\it et
		al.\/} 2018 \mnras\ 477, 5191.
	\bibitem[\protect\citeauthoryear{Rieke {\it et al.\/}}{2005}]{R05}
		Rieke, G. H., Su, K. Y. L., Stansberry, J. A. {\it et al.\/}
		2005 \apj\ 620, 1010.
	\bibitem[\protect\citeauthoryear{Scheeres}{2007}]{S07} Scheeres, D.
		J. 2007 Icarus 189, 370.
	\bibitem[\protect\citeauthoryear{Slack}{1980}]{S80} Slack, G. A.
		1980 \prb\ 22, 3065.
	\bibitem[\protect\citeauthoryear{Spencer {\it et al.\/}}{2020}]{S20}
		Spencer, J. R., Stern, S. A., Moore, J. M. {\it et al.\/}
		2020 Science 367, eaay3999.
	\bibitem[\protect\citeauthoryear{Steckloff {\it et al.\/}}{2021}]
		{St20} Steckloff, J. K., Lisse, C. M., Safrit, T. K., Bosh,
		A. S., Lyra, W. \& Sarid, G. 2021 Icarus 356, 113998
%		\url{https://doi.org/10.1016/j.icarus.2020.113998}
		arXiv:2007.12657.
	\bibitem[\protect\citeauthoryear{Stern}{2003}]{S03} Stern, S. A.
		2003 Nature 424, 639.
	\bibitem[\protect\citeauthoryear{Stern {\it et al.\/}}{2019}]{S19}
		Stern, S. A., Weaver, H. A., Spencer, J. R. {\it et al.\/}
		2019 Science 364, 649.
	\bibitem[\protect\citeauthoryear{Umurhan {\it et al.\/}}{2020}]{U20}
		Umurhan, O. M., Keane, J. T., Beyer, R. A. {\it et al.\/}
		2020 Bull. AAS 52 (\#235), 419.05
%	\bibitem[\protect\citeauthoryear{wikipedia}{2019}]{wikipedia}
%		\url{https://en.wikipedia.org/wiki/(486958)\_2014\_MU69}
%		downloaded Jan. 25, 2019.
\end{thebibliography}
\end{document}